\documentclass[12pt]{article}
\usepackage{amssymb, amsfonts, graphicx, hyperref, psfrag, epsfig, mathrsfs}
\newcommand{\be}{\begin{equation}}
\newcommand{\ee}{\end{equation}}
\newcommand{\bea}{\begin{eqnarray}}
\newcommand{\eea}{\end{eqnarray}}
\newcommand{\ba}{\begin{array}}
\newcommand{\ea}{\end{array}}

\def\Tr{\mathop{\rm Tr}\nolimits}

\csname @addtoreset\endcsname{equation}{section}

\begin{document}

\begin{titlepage}

\title{\bf\Large Note on Chiral Symmetry Breaking from Intersecting Branes\vspace{18pt}}
\author{\normalsize Yi-hong Gao$^a$, Jonathan P.~Shock$^a$, Wei-shui Xu$^a$ and Ding-fang Zeng$^b$ \vspace{12pt}\\
${}^a${\it\small Institute of Theoretical Physics}\\ {\it\small P.O. Box 2735, ~Beijing 100080, P. R.~China}\\
${}^b${\it\small College of Applied Science, Beijing University Of
Technology}\\{\it\small Beijing 100022, P.~R.~China}\\{\small
e-mail: { \it gaoyh@itp.ac.cn,~\it jps@itp.ac.cn,~\it
wsxu@itp.ac.cn,~\it dfzeng@bjut.edu.cn}}}

\date{}
\maketitle  \voffset -.2in \vskip 2cm \centerline{\bf Abstract}
\vskip .4cm In this paper, we will consider the chiral symmetry
breaking in the holographic model constructed from the intersecting
brane configuration, and investigate the Nambu-Goldstone bosons
associated with this symmetry breaking.


\vskip 5.5cm \noindent April 2007 \thispagestyle{empty}
\end{titlepage}


\newpage
\section{Introduction}
In \cite{maldacena1997}, the author proposed that type IIB string
theory on the space $AdS_5\times S^5$ is dual to $\mathcal{N}=4$
supersymmetric gauge theory on the boundary of this geometry, i.e
the AdS/CFT correspondence. Using this method, one can study
strongly coupled physics at zero and finite temperature
\cite{polchinski2000, klebanov2000, maldacena2000, witten1998}.
Since there exist no flavor degrees of freedom in the above
constructions, in \cite{karch2002}, flavor D-brane probes were
introduced into the holographic D-brane constructions in order to
get more realistic models. Along this line, there are many further
developments \cite{kruczenshi2003, sakai2004, antonyan200604,
antonyan200608, gao2006, basu2006} of holographic models with
flavor.

In the framework of these holographic D-brane constructions, many
properties of strongly coupled gauge theory have been investigated.
For example, in \cite{sakai2004, antonyan200604, antonyan200608,
gao2006, basu2006, gepner2006, babington2004, karch2006,
casero2007}, chiral symmetry breaking has be studied in various
dimensional gauge theories. In the confinement phase, such a
symmetry is always broken. But in the deconfinement phase, there
exists a first order phase transition at some critical temperature
$T_{\chi}$, under which the chiral symmetry is broken, while above
this temperature the symmetry is restored \cite{antonyan200608,
gao2006, aharony2006, parnachev2006}.

From field theory, we know that spontaneous breaking of a global
symmetry will give rise to some massless Nambu-Goldstone (NG)
bosons. Hence in the holographic models, there should also exist
some NG bosons associated with the chiral symmetry. Recently, these
NG bosons have been investigated in some models constructed with
D-branes \cite{sakai2004, antonyan200608, gepner2006}. For some
D-brane configurations, for example, the D8/$\overline{D8}$/D4
construction in \cite{sakai2004},  there exists a normalized
massless NG boson associated with the chiral symmetry breaking.
However, for some other configurations, no corresponding normalized
NG bosons exists \cite{antonyan200608, gepner2006}.

In this paper, we follow the holographic D-brane construction in
\cite{gao2006}, which produces the non-local Gross-Neveu model in
the weak coupling regime. In the strong coupling regime, the
supergravity approximation can be used to analyze the underlying
physical system. In this construction, the chiral symmetry will be
spontaneously broken. Here we investigate the massless NG boson
associated with this chiral symmetry breaking, and does not see the
normalized NG boson associated with the $U(1)$ global chiral
symmetry. For the chiral symmetry $U(N_f)_L\times U(N_f)_R$, we
still don't find that the NG bosons will be present in this
holographic construction.

The plan of this paper is as follows. In section 2, we provide a
review of the brane construction in \cite{gao2006}. In section 3, we
study the massless Nambu-Goldstone boson associated with chiral
symmetry breaking. Finally, in the last section, we give some
discussions and conclusions.

\section{A review about brane configuration }

\begin{figure}[ht]
 \centering
 \includegraphics[width=0.55\textwidth]{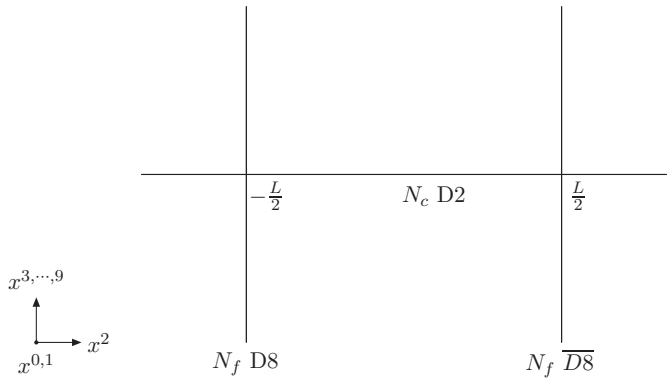}
 \caption{The brane configuration consists of D2, D8 and $\overline{D8}$ branes}
 \label{d2d8}
\end{figure}

In this section, we will give a review of the model \cite{gao2006}
constructed from intersecting D-branes. This brane configuration in
IIA string theory is made up of the $N_c$ D2, $N_f$ D8 and $N_f$
$\overline{D8}$-branes. The extended directions of these branes are
indicated as follows \be
\begin{array}{ccccccccccccc}
 &0 &1 &2 &3 &4 &5 &6 &7 &8 &9\\
 D2: &{\rm x} &{\rm x} &{\rm x} &{} &{} &{} &{} &{} &{} &{}\\
 {D8/\overline{D8}}: &{\rm x} &{\rm x} &{} &{\rm x} &{\rm x} &{\rm x} &{\rm x}
 &{\rm x} &{\rm x} &{\rm
  x}
\end{array}
\label{con} \ee In this brane configuration, the $N_f$ D8 and
$\overline{D8}$-branes are parallel and separated by a distance $L$
in the $x^2$ direction. The $N_c$ D2-branes intersect with the $N_f$
D8 and $\overline{D8}$-branes along the coordinates $(x^0,x^1 )$.
All other coordinates $(x^3,x^4, ... , x^9)$ are transverse
directions to the intersection region of this brane configuration.
The explicit picture of this configuration is shown in fig.
\ref{d2d8}

In this brane configuration, the low energy effective theory is
obtained from the 2-2, 8-8, $\bar{8}$-$\bar{8}$, 8-$\bar{8}$, 2-8
and 2-$\bar{8}$ open strings. In the intersectional dimensions, as
analyzed in \cite{gao2006}, all the massless modes(quarks $q_L$ and
$q_R$, gauge field $A_\mu$) and their transformations under the
gauge symmetries can be listed in the following table \ref{sdfdfs}.
\begin{table}[h] \tabcolsep0.33in
 \centering{
\begin{tabular}{|c|c|c|c|}
 \hline field &SO(1,1) &SO(8) & $U(N_c)\times
U(N_f)_L\times
U(N_f)_R$ \\
\hline   $A_\mu$   &2      &1       &(adj, 1, 1 )\\
\hline
  $q_L$ &$1_+$ &1 & $(N_c, N_f, 1)$\\
  $q_R$ &$1_-$ &1 & $(N_c, 1, N_f)$\\
  \hline
\end{tabular}
}\caption{The massless degrees of freedom on the intersection of
this brane configuration} \label{sdfdfs}
\end{table}

The low energy theory on the worldvolume of the D2-branes is a three
dimensional $U(N_c)$ gauge theory with the gauge coupling constant
$g_3^2 = g_s/l_s$. Under the limit, $g_3^2$ fixed and
$\alpha'\rightarrow 0 $, this theory decouples from the bulk physics
since the ten dimensional Newton constant goes to zero. The 't Hooft
coupling constant can be defined as $\lambda = g_3^2
N_c/4\pi=g_sN_c/4\pi l_s$. Now we analyze the coupling constant
parameter space. As in \cite{itzhaki1998}, we introduce an energy
scale $U\equiv r/\alpha'$, at this energy scale, the effective
dimensionless coupling constant in the three dimensional gauge
theory is $g^2_{eff}\approx\lambda/U$. If $g_{eff}^2\gg 1$, the
theory will be strong coupled, however, on the other side, the
perturbative analysis is valid. Here, the energy scale is the
distance $L$ between D8 and $\bar{D8}$ brane in the unit
$\alpha'=1$. In the regime \be l_s \ll L\ll \frac{1}{\lambda},
\label{weak} \ee since the distance scale $L$ is much larger than
the string scale $l_s$, stringy effects can be neglected. The
effective coupling constant $g^2_{eff}\approx\lambda L\ll 1$, hence
the coupling is weak and the perturbative calculation can be
trusted. We can use the perturbative theory to describe the dynamics
of quarks $q_L$ and $q_R$. In \cite{gao2006}, we find the low energy
effective theory is the non-local Gross-Neveu (GN) model.

If $L\rightarrow \infty$, the $g^2_{eff}\rightarrow \infty$, which
means the interaction between the left-hand quarks and the
right-hand quarks becomes strong with increasing distance $L$.
However, if we let $L\rightarrow 0$, the effective coupling constant
$g^2_{eff}\rightarrow 0$. When the 't Hooft coupling increases into
the regime \be l_s\ll \frac{1}{\lambda} \ll L ,~~~~~
\frac{1}{\lambda} \ll l_s \ll L , \label{strong} \ee the interaction
between $q_L$ and $q_R$ becomes strong. We can't use the above
perturbative method to perform such a calculation, instead in this
regime we can use the SUGRA/Born-Infeld approximation to study the
low energy dynamics of the brane system.

\begin{figure}[h]
\centering{\includegraphics[scale=0.75]{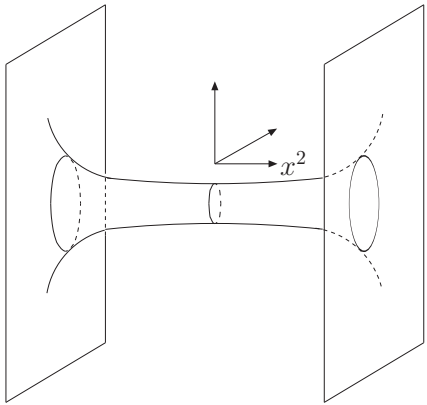}}
 \caption{ The $N_f$ D8 and $\overline{D8}$ branes will be connected in the
 background (\ref{metric}).}
 \label{gn2}
 \end{figure}

The near-horizon geometry of the $N_c$ D2-branes is given by
\bea\ba{lll} &ds^2=\left(\frac{U}{R}\right)^{5/2}
\left(\eta_{\mu\nu} dx^\mu dx^\nu +\left(dx^2\right)^2 \right)
+\left(\frac{R}{U} \right)^{5/2} \left(\left(dU\right)^2
+\left(U\right)^2d\Omega_6^2
\right)\\
&e^\phi=g_s\frac{R^{5/4}}{U^{5/4}}, ~~~~~C_{012}=-\frac{1}{2}
\left(\frac{U^5}{R^5}-1\right),~~~~~ R^5=6\pi^2g_sN_c=6\pi g_3^2N_c
\label{metric}\ea\eea where $\Omega_6$ is the angular direction in
$(3456789)$ and $U= r/\alpha'$ with transverse radial coordinate
$r$. Then we introduce a D8-brane to probe the geometry
(\ref{metric}) (Since the gauge field on the D8 brane isn't turned
on, the results of the $N_f$ coincident D8-branes case is same. In
the next section, we will turn on the fluctuation of the gauge field
on the D8 branes). The embedding of the D8-brane forms a curve
$U=U(x^2)$ in the $(U,x^2)$ plane, whose shape is determined by the
equations of motion that follows from the DBI (Dirac-Born-Infeld)
action. In the background (\ref{metric}), the induced metric on the
D8-brane is \be ds^2=\left(\frac{U}{R}\right)^{5/2}\eta_{\mu\nu}
dx^\mu dx^\nu +\frac{R^{5/2}}{U^{1/2}}d\Omega_6^2
+\left(\left(\frac{U}{R}\right)^{5/2}\left(\frac{\partial
x^2}{\partial U}\right)^2
+\left(\frac{R}{U}\right)^{5/2}\right)dU^2. \label{metric1}\ee The
DBI action for D8-brane is \be S_{D8}\sim\int dx^2
U^{7/2}\sqrt{{1+\left(\frac{R}{U}\right )^5}U'^2},
 \label{act}\ee
where the $U'= dU/dx^2$. From the equation (\ref{act}), the equation
of motion can be obtained\be
\frac{d}{dx^2}\left(\frac{U^{7/2}}{\sqrt{1+\left({\frac{R}{U}}\right)^5
U'^2}}\right)=0.\ee Then we can get the first order differential
equation\be \frac{U^{7/2}}{\sqrt{1+\left({\frac{R}{U}}\right)^5
U'^2}}= U_0^{7/2}.\label{eee} \ee The solution $U(x^2)$ of
(\ref{eee}) is a curve in the (U, $x^2$) plane, which is symmetric
under the reflection $x^2\rightarrow -x^2$. We choose the following
boundary conditions: If the $U \rightarrow \infty$, then
$x^2=\pm\frac{L}{2}$, and at $x^2=0$ the $U$ is equal to $U_0$.
Thus, the solution $x^2(U)$ can be obtained as the integral form \be
x^2(U)=\int_{U_0}^U\frac{dU}{\left(\frac{U}{R}\right)^{5/2}
\sqrt{\left(\frac{U^7}{U_0^7}-1\right)}}.\label{dd}\ee Under the
approximation $U/U_0\gg1$, the curve in the (U, $x^2$) plane can be
obtained \be x^2(U)= \frac{ R^{5/2}}{7U_0^{3/2}}\left(B(\frac{5}{7},
\frac{1}{2})-B(\left(\frac{U_0}{U}\right)^7, \frac{5}{7},
\frac{1}{2})\right).\label{aa}\ee From equation (\ref{aa}), we know
the asymptotic value $ L/2 = x^2(\infty) =\frac{R^{5/2}}{7
U_0^{3/2}}B(\frac{5}{7}, \frac{1}{2}) $. At small $U/U_0$, i.e the
large $U$, the form of curve $U(x^2)$ obeys the equation \be U^5 =
\frac{R^{5/2}U_0^{7/2}}{5\left(L/2-x^2(U)\right)}.\label{ee} \ee
Since the symmetry $x^2\rightarrow -x^2$, the part of the D-brane at
$x^2< 0$ is determined by $U(x^2)= U(-x^2)$. The full D8-brane flow
can be determined in the background (\ref{metric}) and is shown in
fig. \ref{gn2}

In the classical case, we have discussed that the $N_f$ D8 and $N_f$
$\overline{D8}$-branes in this brane configuration sit at $x^2
=-\frac{L}{2}$ and $x^2 = \frac{L}{2}$ respectively, and are
separated by a distance $L$. Hence the chiral symmetry
$U(N_f)_L\times U(N_f)_R$ is not broken. But due to the quantum
effects, the $N_f$ D8-branes and $N_f$ $\overline{D8}$-branes are
joined into a single D8-branes by a wormhole and the chiral symmetry
$U(N_f)_L \times U(N_f)_R$ is dynamically broken to a
$U(N_f)_{diag}$. We can compare the energy density of these two
configuration to see which one is preferred. The energy density
difference of the two configurations is given by \bea
&&\bigtriangleup E \equiv E_{straight}-E_{curved} \nonumber \\
&&\sim \int_0^{U_0} (U - 0)dU + \int_{U_0}^\infty
\left(U-U(1-\frac{U_0^7}{U^7})^{-1/2}\right)dU \approx 0.4 U_0^2.
\label{difference}\eea We find the energy density difference
$\bigtriangleup E$ is positive. This result means that the
configuration connected by the wormhole is preferred and the chiral
symmetry is broken.

In order for the supergravity method to be reliable in this regime,
two conditions must be satisfied. One is $e^\phi \ll 1$, and the
other is that scalar curvature satisfies $\alpha'R\ll 1$. Therefore
the 't Hooft coupling constant must be satisfied \be U \ll \lambda
\label{bc}\ee which is extended by the eleven-dimensional SUGRA.
From (\ref{aa}), we have $U_0^3 \sim \lambda / L^2$. Actually, $U_0$
obviously satisfies the condition (\ref{bc}). Fixing $L$ and
increasing $\lambda$ will push $U_0$ further into the region of
validity of supergravity. In the opposite direction, decreasing
$\lambda$ will make $U_0$ become smaller, and when $L\simeq
\frac{1}{\lambda}$ the curvature at $U_0$ becomes of order one and
the supergravity description breaks down. By continuing to decrease
$\lambda$ into the region $L \ll \frac{1}{\lambda}$, the coupling
becomes weak and the perturbative description is valid.

\section{Nambu-Goldstone boson }
In the last section, we have given a review of the chiral symmetry
breaking in the holographic model from the intersecting D-brane
configuration D8/$\overline{D8}$/D2. Now, in the following, we
investigate the Nambu-Goldstone bosons associated with this chiral
symmetry breaking. Following the method in \cite{sakai2004,
antonyan200608, gepner2006}, we can turn on the fluctuations of the
gauge field on the worldvolume of the flavor D8-branes. The gauge
field on this probe D8-brane is labeled by $A_M$,
($M=0,~1,~U,~d\Omega_6$)\footnote{Here we only consider the
coordinate $U$ as a function of the coordinate $x^2<0$, for the case
$x^2>0$, which can be simply obtained by the symmetry
$U(x^2)=U(-x^2)$. Thus, in the following, we can take the gauge
field on the D8 brane as a single-valued function of the coordinate
$U$.}. Since we are mainly interested in the $SO(7)$ singlet states,
we can set the components $A_{\alpha}$ of the gauge field on the
$S^6$ to vanish and the other components $A_{\mu} (\mu=0, 1)$, $A_U$
to be independent of the coordinates on the sphere $S^6$. Then, for
the $U(1)$ case, the effective action of the
D8-brane in the background (\ref{metric}) is \bea S&=&S_{DBI}+S_{cs}\nonumber \\
S_{DBI}&=&-T_8\int
d^9xe^{-\phi}\sqrt{-\det(g_{MN}+2\pi F_{MN})} \nonumber \\
S_{cs}&=&T_8\int (C_3+C_5)\wedge e^{2\pi F_2}.\eea The first term in
the CS (Chern-Simons) terms can be omitted due to the three powers
of the gauge field strength $F_2$. However, to the second term,
which has the same order as the DBI action part, hence, we can throw
it away. We choose that the Hodge dual of the RR (Ramond-Ramond)
field $F_4$ is $F_6=\frac{N_c}{V_{S^6}}\epsilon_{6}$, where the
$V_{S^6}$, $\epsilon_{6}$ are the volume of the sphere $S^6$ volume
form of the sphere $S^6$ respectively. Thus, the contribution from
the CS-term is \bea S_{cs}&=&4\pi^2T_8\int C_5\wedge F_2\wedge
F_2+O(F_2^3)\nonumber\\
&=&4\pi^2N_cT_8\int A\wedge F_2+O(F_2^3)\eea

After expanding the field strength and omitting the high order
terms, we can obtain the action \be S=-(2\pi)^2T_8V_{S^6}\int d^2xdU
e^{-\phi}\sqrt{-\det
g_{MN}}[\frac{1}{4}F_{MN}F^{MN}]+4\pi^2N_cT_8\int A\wedge F_2. \ee
Now we can substitute the induced metric of the D8-brane into the
equation (\ref{eee}), the above equation becomes \bea
S=-\frac{(2\pi)^2T_8V_{S^6}}{g_s}\int
d^2xdU\left[\frac{1}{4}G(U)F_{\mu\nu}F^{\mu\nu}+\frac{1}{2}H(U)F_{\mu
U}F^{\mu}_U\right]\nonumber\\+2\pi^2N_cT_8\int
d^2xdU\epsilon^{\mu\nu}\left[A_UF_{\mu\nu}+2A_\mu F_{\nu U}\right]
\label{action} \eea where the indices are contracted under the
Minkowski metric $\eta_{\mu\nu}$, and the $G(U)$, $H(U)$ are defined
in the following equations \bea
G(U)&=&\frac{R^{10}}{U^{1/2}\sqrt{U^7-U_0^7}},\nonumber\\
H(U)&=&\frac{R^5\sqrt{U^7-U_0^7}}{U^{5/2}}.\eea

We can expand the gauge field components $A_\mu$ and $A_U$ in terms
of the complete basis $\psi_n(U)$ and $\phi_n(U)$ as follows \bea
A_\mu(x,U)=\sum_{n} B_\mu^n(x)\psi_n(U),\nonumber\\
A_U(x,U)=\sum_{n} \varphi^n(x)\phi_n(U). \eea Then the gauge field
strength will be \bea F_{\mu\nu}(x,U)&=&\sum_n \left(\partial_\mu
B^n_\nu(x)
-\partial_\nu B^n_\mu(x)\right)\psi_n(U)\nonumber\\
&\equiv&\sum_{n=0}^{\infty} B_{\mu\nu}^n(x)\psi_n(U),\\
F_{\mu U}(x,U)&=&\sum_n \left(\partial_\mu \varphi^n(x)\phi_n(U)
-B^n_\mu(x)\dot\psi_n(U)\right)\eea where the $\dot\psi_n(U)$
denotes the $\partial_U\psi_n(U)$. Inserting the above two equations
into the action (\ref{action}), we get \bea S&=&-
\frac{(2\pi)^2T_8V_{S^6}}{g_s}\int d^2x dU \sum_{m,n} \left[
\frac{1}{4}G(U) B_{\mu\nu}^mB^{n\mu\nu }\psi_m\psi_n \right.
\nonumber \\ &&\left. + \frac{1}{2}H(U) \left(\partial_\mu
\varphi^m\partial^\mu \varphi^n\phi_m\phi_n +B_\mu^m
B^{n\mu}\dot\psi_m\dot\psi_n
-2\partial_\mu\varphi^mB^{n\mu}\phi_m\dot\psi_n \right)
\right]\nonumber\\ &&+2\pi^2N_cT_8\int
d^2xdU\epsilon^{\mu\nu}\sum_{m,n}\left[B^m_{\mu\nu}\varphi^n\psi_m\phi_n
\right. \nonumber \\ &&\left. +
2(B^m_\mu\partial_\nu\varphi^n\psi_m\phi_n-B^m_\mu
B^n_\nu\psi_m\dot\psi_n)\right]. \label{action1}\eea We set the
basis $\psi_n(U)$ to satisfy the following normalization condition
\be \frac{(2\pi)^2T_8V_{S^6}}{g_s}\int dU
G(U)\psi_n(U)\psi_m(U)=\delta_{mn}. \ee Then the first term in the
equation (\ref{action1}) will become \be -\sum_{n=1}\int d^2x
\frac{1}{4} B_{\mu\nu}^nB^{n\mu\nu} \label{kinetic}\ee which are the
kinetic terms for the gauge field $B_\mu^n$ in two dimensions. If we
choose the field $\psi_n(U)$ $(n\geq 1)$ to satisfy the equation \be
\frac{1}{G(U)}\partial_U[H(U)\dot\psi_n(U)]=-m_n^2\psi_n(U),\ee then
$\dot\psi_n(U)$ satisfies the normalization condition \be
\frac{(2\pi)^2T_8V_{S^6}}{g_s}\int dU
H(U)\dot\psi_n(U)\dot\psi_m(U)=m_n^2\delta_{mn}.
\label{masscondition}\ee From the equation (\ref{action1}), we
obtain the mass term for the gauge fields $B_\mu^n$, it is \be
-\sum_{n=1}\int d^2x \frac{1}{2}m_n^2B_\mu^nB^{n\mu}\label{mass} \ee
Thus, for the fields $B_\mu^n$ $(n\geq 1)$, summing the equation
(\ref{kinetic}) and (\ref{mass}), we get the action for these
massive gauge fields in two dimensions \be S_B = -\sum_{n=1}\int
d^2x \left[\frac{1}{4}
B_{\mu\nu}^nB^{n\mu\nu}+\frac{1}{2}m_n^2B_\mu^nB^{n\mu}\right].\ee

For the complete basis $\phi_n(U)$, we impose the normalization
condition \be \frac{(2\pi)^2T_8V_{S^6}}{g_s}\int dU
H(U)\phi_n(U)\phi_m(U)=\delta_{mn}. \ee From the equation
(\ref{masscondition}), we let $\phi_n=m_n^{-1}\dot\psi_n$ for the
$n\geq 1$ cases. For the zero mode $\phi_0$, if we choose
$\phi_0=C/H(U)$, then \be \frac{(2\pi)^2T_8V_{S^6}}{g_s}\int dU
H(U)\phi_0(U)\phi_m(U)=\frac{(2\pi)^2T_8V_{S^6}}{m_ng_s}\int dU
H(U)\phi_0\dot\psi_m\sim \int dU\dot\psi_n=0.\ee Hence the zero mode
$\phi_0$ is orthogonal to the basis $\phi_n$ and $\dot\psi_n$ for
all $n\geq 1$. The normalization condition  of the zero mode
$\phi_0(U)$ is \be \frac{(2\pi)^2T_8V_{S^6}}{g_s}\int dU
H(U)\phi_0(U)\phi_0(U)=\frac{(2\pi)^2T_8V_{S^6}C^2}{g_s}\int dU
H^{-1}.\ee Due to the $H=\frac{R^5\sqrt{U^7-U_0^7}}{U^{5/2}}$, then
the integral $\int dU H^{-1}$ will be logarithmic divergence. It
means that the zero mode $\phi_0$ can't be normalized. While, due to
the integral $\int dU G(U)\psi_0\psi_0$ is convergent, another zero
constant mode $\psi_0(U)$ is normalized. All these results are same
as the corresponding ones in \cite{antonyan200608}, but are
different from the ones in \cite{sakai2004}.

And using the definition \be  2\pi^2N_cT_8\int
dU\psi_m\dot\psi_n=M_{mn}, ~~2\pi^2N_cT_8\int
dU\phi_0\psi_n=M_m.\label{condition2}\ee Then the full fluctuation
action is \bea S&=& -\int d^2x\left[\frac{1}{4}
B_{\mu\nu}^0B^{0\mu\nu}+\sum_{n=1}\left(\frac{1}{2}\partial_\mu\varphi^n\partial^\mu\varphi^n+\frac{1}{4}
B_{\mu\nu}^nB^{n\mu\nu}+\frac{1}{2}m_n^2B_\mu^nB^{n\mu}\right.\right. \nonumber \\
&&\left.\left.-m_n\partial_\mu\varphi^nB^{n\mu}\right)\right]+\sum_{m,n=1}\int
d^2x\epsilon^{\mu\nu}\left[M_m(\varphi^0B^m_{\mu\nu}+2B^m_\mu\partial_\nu\varphi^0)
\right. \nonumber \\ &&\left.
+M_{mn}(m_n^{-1}B^m_{\mu\nu}\varphi^n-2B^m_\mu(B^n_\nu-m_n^{-1}\partial_\nu\varphi^m))\right].\label{dd}\eea
Through the gauge transformation \be B^n_\mu\rightarrow
B^n_\mu-m_n^{-1}\partial_\mu \varphi^n, \label{transformation}\ee
the $\partial_\mu\varphi^n$ can be absorbed into the field $B_\mu^n$
in the first line of the equation (\ref{dd}). Hence the final action
 is \bea S&=&-\int d^2x \left[\frac{1}{4} B_{\mu\nu}^0B^{0\mu\nu}
+ \sum_{n=1}\left( \frac{1}{4} B_{\mu\nu}^nB^{n\mu\nu} +\frac{1}{2}
m_n^2B_\mu^nB^{n\mu}\right)\right]\nonumber\\ &&+\sum_{m,n=1}\int
d^2x\epsilon^{\mu\nu}\left[M_n(\varphi^0B^n_{\mu\nu}+2(B^n_\mu+m_n^{-1}\partial_\mu\varphi^n)
\partial_\nu\varphi^0)-2M_{mn}B^m_\mu B^n_\nu)\right].
 \label{fluc1}\eea In the above equation, there exists some coupling terms between the zero mode $\varphi_0$
and other modes $
\epsilon^{\mu\nu}M_n[\varphi^0B^n_{\mu\nu}+2(B^n_\mu+m_n^{-1}\partial_\mu\varphi^n)
\partial_\nu\varphi^0]$. And since the $\phi_0(U)$ is not normalized,
the equation (\ref{fluc1}) doesn't have the kinetic term of the mode
$\varphi^0$. Thus, we can't regard the $\varphi^0$ as a massless
field, and can't be taken as the Nambu-Goldstone boson associated
with the chiral symmetry breaking.

As in \cite{sakai2004}, we can change inot the $A_U=0$ gauge, which
can be chosen due to the gauge transformation \be A_M\rightarrow
A_M-\partial_M\Lambda \ee with the
$\Lambda=\sum_{n=1}m_n^{-1}\varphi^n\psi_n(U)$. Then after
substituting these into the equation (\ref{action}), we can get the
action is \bea S&=&-\int
d^2x\left[\frac{1}{4}B^0_{\mu\nu}B^{0\mu\nu}+\sum_{n=1}\left(\frac{1}{4}B^n_{\mu\nu}B^{n\mu\nu}
+\frac{1}{2}m_n^2B^n_\mu B^{n\nu}\right)\right]\nonumber\\&& -2\int
d^2x \sum_{m,n}\epsilon^{\mu\nu}M_{mn}B^n_\mu B^m_\nu. \eea  This
action is same as the equation (\ref{fluc1}) after throwing out the
zero mode $\varphi_0$ due to the non-normalization of $\phi_0$. And
this result is also same as the one in the \cite{antonyan200608,
gepner2006}. As the same arguments in \cite{antonyan200608}, it is
difficult to diagonalize the infinite-dimensional matrix, but
generally the mass eigenvalues of the meson fields $B^n_\mu$ does
not vanish.

Thus, for the D8/$\overline{D8}/D2$ brane system, after the above
analysis, we doesn't find the NG-boson associated with the $U(1)$
chiral symmetry breaking. The reason may be the NG boson will not be
visible in the analysis of the near horizon geometry because these
degrees lives a far distance from the D2 brane
\cite{antonyan200608}.

In order to investigate the chiral symmetry $U(N_f)_L\times
U(N_f)_R$ broken to $U(N_f)_{diag}$, we need generalize to the $N_f$
flavor D8 branes case. For the multi-flavors to probe the near
horizon background (\ref{metric}), we need use the non-Abelian DBI
action to describe the dynamics of the D8 branes \cite{myers1999}.
Using the same ansantz for the gauge field as the $U(1)$ case and
omitting the higher order terms of the field strength, the action of
the gauge field on the D8-branes in the background of solution
(\ref{aa}) reads \bea S&=&-\frac{(2\pi)^2T_8V_{S^6}}{g_s}\int d^2x
dU \Tr\left[\frac{1}{4}G(U) F_{\mu \nu}F^{\mu\nu} +\frac{1}{2}H(U)
F_{\mu U}F^\mu_U \right]\nonumber\\ && +4\pi^2N_cT_8 \Tr\int
\left(A\wedge dA+\frac{2}{3}A\wedge A\wedge A\right)\label{multi}
\eea where the field strength is $F_{\mu\nu}=\partial_\mu
A_\nu-\partial_\nu A_\mu+[A_\mu, A_\nu]$, and the trace
$\Tr{T^aT^b}=N_f\delta^{ab}$ under the gauge group $U(N_f)$.

We then expand the gauge field in the complete basis $\psi_n(U)$ and
$\phi_n(U)$ as the same in the $U(1)$ case, except where the modes
$B_\mu^n(x)$ and $\varphi^n(x)$ transform under the adjoint
representation of the gauge group $U(N_f)$. Then using the
normalization conditions as same in the $U(1)$ case and the
following definitions \bea &&\frac{(2\pi)^2T_8V_{S^6}}{g_s}\int dU
G(U)\psi_n(U)\psi_m(U)\psi_k(U)=A_{nmk},\label{c1}\\
&& \frac{(2\pi)^2T_8V_{S^6}}{g_s}\int dU
G(U)\psi_n(U)\psi_m(U)\psi_k(U)\psi_l(U)=B_{nmkl},\label{c2}\\
&&\frac{(2\pi)^2T_8V_{S^6}}{g_s}\int dU
H(U)\psi_n(U)\phi_m(U)\phi_k(U)=C_{nmk},\label{c3}\\
&&\frac{(2\pi)^2T_8V_{S^6}}{g_s}\int dU
H(U)\psi_n(U)\dot\psi_m(U)\phi_k(U)=D_{nmk},\label{c4}\\
&&\frac{(2\pi)^2T_8V_{S^6}}{g_s}\int dU
H(U)\psi_n(U)\psi_m(U)\phi_k(U)\phi_l(U)=E_{nmkl}\label{c5}, \eea we
can get all the terms of the $F_{\mu\nu}F^{\mu\nu}$ and $F_{\mu
U}F^{\mu U}$ in the first line of the action (\ref{multi}) as
follows\bea
&&-\frac{1}{4}\Tr\left[\sum_{n=0}B^n_{\mu\nu}B^{n\mu\nu}+ 2\sum_{n,
m,k=0}A_{nmk}B^n_{\mu\nu}[B^{m\mu}, B^{k\nu }]\right. \nonumber\\
&&\left.
~~~~+ \sum_{n, m, k, l=0}B_{nmkl}[B_\mu^m, B_\nu^n][B^{k\mu}, B^{l\nu}]\right]\nonumber\\
&&-\frac{1}{2}\Tr\left[\sum_{n=1}\left(\partial_\mu\varphi^n\partial^\mu\varphi^n
-2m_n\partial_\mu\varphi^nB^{n\mu}+m_n^2B^n_\mu
B^{n\mu}\right)\right.\nonumber\\ &&\left. +2\sum_{n, k\geq1;
m=0}\left(C_{mnk}\partial_\mu\varphi^n-D_{mnk}B_\mu^n\right)
[B^{m\mu}, \varphi^k]\right.\nonumber\\  &&\left.~~~~~~+\sum_{n,
l\geq1; m, k=0}E_{mknl}[B^m_\mu, \varphi^n][B^{k\mu},
\varphi^l]\right] \label{fluc2}\eea where the
$B^n_{\mu\nu}=\partial_\mu B^n_\mu-\partial_\nu B^n_\mu$. For the
$N_f=1$ case, since the constants $A_{nmk}$, $B_{nmkl}$, $C_{nmk}$,
$D_{nmk}$, $E_{nmkl}$ all vanish, the above equations will reduce to
equation (\ref{fluc1}) through the gauge transformation
(\ref{transformation}). However, if $N_f\neq 1$, the constants
$A_{nmk}, B_{nmkl}, C_{nmk}, D_{nmk}$ and $E_{nmkl}$ cannot vanish
all together.

The second line, setting to be $\chi$, in the action (\ref{multi})
contributed from the CS term, after substituting the expansion of
the gauge field, reads \bea \chi &=&2\pi^2N_cT_8\Tr\int d^2xdU
\epsilon^{\mu\nu}\left( A_U(\partial_\mu A_\nu-\partial_\nu
A_\mu)+\frac{4}{3}A_\mu A_\nu
A_U\right)\nonumber\\&=&2\pi^2N_cT_8\Tr\int d^2xdU
\epsilon^{\mu\nu}\sum_{m,n,k}\left[B^m_{\mu\nu}\varphi^n\psi_m\phi_n
\right. \nonumber \\ &&\left.+
2(B^m_\mu\partial_\nu\varphi^n\psi_m\phi_n-B^m_\mu
B^n_\nu\psi_m\dot\psi_n)+\frac{4}{3}B^n_\mu
B^m_\nu\varphi^k\psi_n\psi_m\phi_k\right].\eea Using the same
definition as the equation (\ref{condition2}), and the condition
(\ref{c5}), we can get \bea \chi &= &\sum_{m,n=1}\Tr\int
d^2x\epsilon^{\mu\nu}\left[M_m(\varphi^0B^m_{\mu\nu}+2B^m_\mu\partial_\nu\varphi^0)
\right. \nonumber \\&&\left.
+M_{mn}(m_n^{-1}B^m_{\mu\nu}\varphi^n-2B^m_\mu(B^n_\nu-m_n^{-1}\partial_\nu\varphi^m))\right]
\nonumber\\&&+\frac{2N_cg_sC}{3V_{S^6}}\sum_{n,m,k=0}\Tr\int d^2x
E_{nmk0}\epsilon^{\mu\nu}B^n_\mu B^m_\nu\varphi^k. \label{cs}\eea

From the equation (\ref{fluc2}) and (\ref{cs}), we can see there
doesn't exists the kinetic term of the zero mode $\varphi_0$ due to
the non-normalization, and the modes $\varphi^0$ are not massless NG
bosons. Thus, through the analysis of the gauge field fluctuation on
the $N_f$ D8 branes, we don't find the $N_f^2$ massless NG bosons in
the spectrum corresponding to this chiral symmetry breaking.

\section{Conclusions}
In \cite{gao2006}, the intersecting brane configuration
$D8/\overline{D8}/D2$ was constructed in IIA string theory. The low
energy theory on this brane configuration can be analyzed using two
methods. In the weak coupling regime, the perturbative method is
reliable and the low energy theory is a nonlocal generalization of
the GN model which dynamically breaks the chiral flavor symmetry
$U(N_f)_L \times U(N_f)_R$ at large $N_c$ and finite $N_f$. However,
in the strong coupling region, we can use the supergravity
approximation to describe the low energy dynamics of the brane
system. In the near horizon geometry of $N_c$ D2 branes, we find
that the connected shape of $N_f$ D8 and $\overline{D8}$ through a
throat in fig. \ref{gn2} is preferred to the separated case of $N_f$
D8 and $\overline{D8}$ in fig. \ref{d2d8} from equation
(\ref{difference}). In the connected case of $N_f$ D8 and
$\overline{D8}$ branes, the chiral symmetry $U(N_f)_L\times
U(N_f)_R$ is broken to the gauge group $U(N_f)_{diag}$. Thus,
totally $N_f^2$ generators of the symmetry $U(N_f)_L\times U(N_f)_R$
are broken in this process.

Associated with this global symmetry breaking, there must exist some
massless Nambu-Goldstone bosons in the spectrum. In the section 3,
we have given a detailed analysis of the fluctuation of the gauge
field on the flavor D8 branes. For the $U(1)$ case, since the zero
mode $\phi_0(U)$ is not normalized, we can't find one massless
Nambu-Goldstone boson in the spectrum which is corresponding to the
chiral symmetry breaking. For the $U(N_f)$ case, we already know
that the chiral symmetry $U(N_f)_L\times U(N_f)_R$ is be broken to
$U(N_f)_{diag}$ in section 2.  However, we still don't see the NG
modes in the spectrum with the same reason as in the $U(1)$ case. So
the results in this two dimensional model are different from the
ones in \cite{sakai2004}, but are consistent with
\cite{antonyan200608, gepner2006, Grisa:2006tm}.

It may be interesting to generalize to other holographic models,
constructed from brane configurations such as Dp/$\overline{Dp}$/D2
$(p=4, 6)$. The intersecting region of these brane configurations is
still two dimensional, $(x^0, x^1)$. In these intersecting
dimensions, in the weak coupling regime, the low energy physics can
be described by the effective field theory. In the strong coupling
regime, the supergravity method can be used to analyze the physics
as in the D8/$\overline{D8}$/D2 brane configuration \cite{gao2006}.
In the near horizon geometry of $N_c$ D2 branes, we find that the
flavor Dp and $\overline{Dp}$ $(p=4, 6)$ branes connect at some
critical point, which means the chiral symmetry is broken. Thus, for
these holographic brane models, one can use the methods in this
paper to analyze the chiral symmetry breaking pattern, and to see
whether the NG modes exist.

\section*{Acknowledgements} We would like to thank Professor Miao li for the
useful discussions, and Professor S. Sugimoto for the
correspondence.


\end{document}